\documentclass[prd,preprint]{revtex4}
\usepackage{amssymb}

\begin{document}

\title{Fermion Propagators in Type II Fivebrane Backgrounds}

\author{Noriaki Kitazawa}
\email{kitazawa@phys.metro-u.ac.jp}
\affiliation{
Department of Physics, Tokyo Metropolitan University,
Hachioji, Tokyo 192-0397, Japan}

\date{\today}

\begin{abstract}
The fermion propagators in the fivebrane background
 of type II superstring theories are calculated.
The propagator can be obtained
 by explicitly evaluating the transition amplitude
 between two specific NS-R boundary states by the propagator operator
 in the non-trivial world-sheet conformal field theory
 for the fivebrane background.
The propagator in the field theory limit
 can be obtained by using point boundary states.
We can explicitly investigate
 the lowest lying fermion states
 propagating in the non-trivial ten-dimensional space-time
 of the fivebrane background:
 $M^6 \times W_k^{(4)}$,
 where $W_k^{(4)}$ is the group manifold of SU$(2)_k \times$U$(1)$.
The half of the original supersymmetry is spontaneously broken,
 and the space-time Lorentz symmetry SO$(9,1)$ reduces to
 SO$(5,1)$ in SO$(5,1) \times$SO$(4) \subset$ SO$(9,1)$
 by the fivebrane background.
We find that there are no propagations
 of SO$(4)$ (local Lorentz) spinor fields,
 which is consistent with the arguments
 on the fermion zero-modes in the fivebrane background
 of low-energy type II supergravity theories.
\end{abstract}

\pacs{}
\preprint{}

\vspace*{3cm}

\maketitle

\section{Introduction}
\label{sec:intro}

The quantum effect of gravity
 may shed new light on the unsolved problems in particle physics.
It was pointed out that
 supersymmetry can be spontaneously and dynamically broken
 by the gravitino condensation\cite{Witten,KMP}
 in non-trivial space-time backgrounds.
It was also suggested that
 all the global symmetries in the low-energy effective field theory
 should be broken by the effect of quantum gravity
 \cite{Giddings-Strominger,Coleman}.
It must be very important
 to investigate these possibilities in the string theory
 which is a strong candidate for the consistent theory
 of quantum gravity.

If we have a fermion propagator, or a fermion two point function,
 in some non-trivial backgrounds (space-time metric or gauge fields),
 we can obtain a value of the fermion pair condensate
 which can be an order parameter of some symmetry breaking.
For example,
 in case of the SU$(N)$ gauge theory
 with massless vector-like fermions
 in the fundamental representation of the gauge group
 in Euclidean space-time,
 the zero-instanton sector in the path integral
 of the fermion two-point function gives Euclidean propagator,
 and the one-instanton sector gives the fermion pair condensate
 which triggers chiral symmetry breaking.
In the same system in Minkowski space-time
 full propagator should include the mass function
 whose integral gives the value of the fermion pair condensate
 through the arguments of the operator product expansion.
Therefore,
 it is interesting to investigate the fermion propagator
 in non-trivial space-time backgrounds
 in string theories.

The fivebrane background
 is a non-trivial background of space-time metric and fields
 in type IIA, type IIB and heterotic string theories
 \cite{Rey1,CHS1,CHS2}.
The half of the original supersymmetry is broken,
 and the space-time Lorentz symmetry reduces to SO$(5,1)$
 from SO$(9,1) \supset$ SO$(5,1) \times$SO$(4)$
 by the fivebrane background.
Although
 general fivebrane backgrounds are realized as the solutions
 in the low-energy supergravity theory of each string theory,
 in some special case they can be described
 by world-sheet conformal field theories
 as the solutions of string theories\cite{Rey2,AFK}.
Therefore, in principle
 we can calculate fermion propagators in string theories
 perturbatively with respect to the string coupling
 without any low-energy approximation
 \cite{Fainberg-Marshakov,Marshakov}.

The number of fermion zero-modes
 in non-trivial four-dimensional space of fivebrane backgrounds
 in low-energy supergravity theories
 is well-known\cite{CHS2}.
There are
 four fermion zero-modes in type II theories\cite{Kitazawa1},
 and two fermion zero-modes in heterotic theories
 \cite{Katagiri-Kitazawa}.
If we believe
 the path integral formalism of the supergravity theory,
 these numbers mean
 four fermion condensations in type II theories
 and fermion pair condensations in heterotic theories
 in non-trivial four-dimensional space.
Although
 the case of the heterotic theory
 is much more interesting than the case of the type II theory
 in the strategy of extracting the value of
 the fermion pair condensate from the propagator,
 there are some technical difficulties to calculate
 the fermion propagator (especially for gravitino and dilatino)
 in heterotic string theories.
 (The calculation of the gaugino propagator is possible,
  and the work is underway.) 
In this paper
 we calculate the gravitino and dilatino propagator
 in the fivebrane background in type II string theories
 as the first attempt.
If the path integral formalism of the supergravity theory is correct,
 there should be no propagations of the light fields
 in SO$(4)$ (local Lorentz) spinor representations.

The paper is organized as follows.
In the next section
 we give a brief review of the fivebrane background
 in type II string theories.
The conformal field theory for the fivebrane background is introduced.
In Sec.\ref{sec:boundary}
 the closed string boundary state
 to which a single fermion state (NS-R state) can couple is introduced.
In Sec.\ref{sec:propagator}
 fermion propagators are calculated by evaluating
 the transition amplitude between boundary states
 by the propagator operator in the world-sheet conformal field theory.
In the last section
 we summarize our results and give some comments.

\section{Fivebrane backgrounds in type II string theories}
\label{sec:CFT}

The fivebrane background (or the NS5-brane) is a BPS configuration
 which preserves half of the supersymmetry of the theory.
In type II supergravity theories,
 which are the low-energy effective theories
 of type IIA and type IIB  string theories,
 the space-time metric configuration of the fivebrane background
 (NS5-branes at the origin)
 is explicitly given by
\begin{eqnarray}
 g_{\mu\nu} &=& e^{2\Phi} \delta_{\mu\nu},
\\
 e^{2\Phi} &=& e^{2\Phi_0} + {{n \alpha'} \over {r^2}},
\label{dilaton}
\end{eqnarray}
where $n$ is an integer,
 $\mu=6,7,8,9$ are the index
 of the non-trivial four-dimensional space
 in the whole ten-dimensional space-time,
 and $r^2=\displaystyle{\sum_{\mu=6}^9 x^\mu x^\mu}$.
We use ${\hat \mu}=0,1,\cdots,9$
 as the index of the whole ten-dimensional space-time,
 and use ${\bar \mu}=0,1,\cdots,5$
 as the index of the flat six-dimensional Minkowski space-time
 in the whole ten-dimensional space-time from now on.
The geometry of the space-time is $M^6 \times R \times S^3$
 with varying radius of $S^3$ from $\sqrt{n\alpha'}$ to infinity
 along with the value of the coordinate of $R$
 from $-\infty$ to $\infty$.
These solutions in low-energy supergravity theories
 are considered to be the exact solutions
 in type II string theories.
The world-sheet conformal field theory for these backgrounds
 can be explicitly constructed in case of $e^{2\Phi_0}=0$.
In this case
 the space-time geometry is $M^6 \times W^{(4)}_k$,
 where $W^{(4)}_k$ is the four-dimensional group manifold
 of SU$(2)_k \times$U$(1)$ with Ka\v c-Moody level $k$.
The geometry of $W^{(4)}_k$ is again $R \times S^3$,
 but the radius of $S^3$ is fixed with $\alpha'/Q$,
 where $Q = \sqrt{\alpha'/(k+2)}$.

In case of the flat space-time
 the world-sheet theory of the type II string theory
 consists of one free bosonic field $X^{\hat \mu}(z,{\bar z})$
 and two free fermionic fields
 $\psi^{\hat \mu}(z)$ and ${\tilde \psi}^{\hat \mu}({\bar z})$.
The system has $(N,{\tilde N})=(1,1)$ superconformal symmetry
 with the central charge $c_m={\tilde c}_m=10+10/2=15$
 which is cancelled by the ghost contribution
 $c_g={\tilde c}_g=-26+11=-15$.
The non-trivial background $M^6 \times W^{(4)}_k$
 can be described by replacing unconstrained fields
 $X^\mu$ ($\mu=6,7,8,9$)
 with the fields constrained on the group manifold of
 SU$(2)_k \times$U$(1)$.
Namely,
 the part of the world-sheet theory
 corresponding to the space-time coordinates of $\mu=6,7,8,9$
 is replaced by the combination of the SU$(2)_k$ WZW model
 and the linear-dilaton theory.
The new part has $(N,{\tilde N})=(4,4)$ superconformal symmetry,
 and has the same central charge of the original part:
 $c_{4D}=6$.

The holomorphic sector of the SU$(2)_k \times$U$(1)$ part
 is described by three SU$(2)_k$ bosonic currents $J_i(z)$
 ($i = 1,2,3$), one free bosonic current $J_4(z)=\partial X^{\mu=6}$
 and four free fermionic fields $\Psi_a$ ($a=1,2,3,4$).
These currents and fields satisfy
 the following operator product expansion.
(We set $\alpha'=2$ from now on in this section for simplicity.)
\begin{eqnarray}
 J_i(z) J_j(z') &\sim&
  - { k \over 2} {{\delta_{ij}} \over {(z-z')^2}}
  + \epsilon_{ijl} {{J_l} \over {z-z'}},
\\
 J_4(z) J_4(z') &\sim& - {1 \over {(z-z')^2}},
\\
 \Psi_a(z) \Psi_b(z') &\sim& - {{\delta_{ab}} \over {z-z'}}.
\end{eqnarray}
The $N=4$ superconformal symmetry transformation is generated
 by the following energy-momentum tensor $T^{W^{(4)}_k}(z)$,
 supercurrents $G^{W^{(4)}_k}_a(z)$ and SU$(2)_n$ currents $S_i(z)$.
\begin{eqnarray}
 T^{W^{(4)}_k} &=&
  - {1 \over 2}
  \left(
   {2 \over {k+2}} J_i^2 + J_4^2
   - \Psi_a \partial \Psi_a + Q \partial J_4
  \right),
\\
 G^{W^{(4)}_k}_i &=&
  \sqrt{{2 \over {k+2}}}
   \left(
    J_i \Psi_4 - \epsilon_{ijl} J_j \Psi_l
    + {1 \over 2} \epsilon_{ijl} \Psi_4 \Psi_j \Psi_l
   \right)
  - J_4 \Psi_i - Q \partial \Psi_i,
\\
 G^{W^{(4)}_k}_4 &=&
  \sqrt{{2 \over {k+2}}}
   \left(
    J_i \Psi_i + {1 \over {3!}} \epsilon_{ijl} \Psi_i \Psi_j \Psi_l
   \right)
  + J_4 \Psi_4 + Q \partial \Psi_4,
\\
 S_i &=&
  {1 \over 2}
  \left(
   \Psi_4 \Psi_i + {1 \over 2} \epsilon_{ijl} \Psi_j \Psi_l
  \right).
\end{eqnarray}
The background charge $Q$
 determines the gradient of the linear-dilaton background
 $\Phi=QX^{\mu=6}$.
The world-sheet field $X^{\mu=6}(z,{\bar z})$
 is called the Feigin-Fuchs field,
 and the theory of this field is called the linear-dilaton theory.
The value $Q=\sqrt{\alpha'/(k+2)}=\sqrt{2/(k+2)}$ is required
 for the correct central charge of $c_{4D}=6$,
 and the Ka\v c-Moody level $n$ of SU$(2)_n$ is fixed to unity
 due to the relation of $n=c_{4D}/6$.
The anti-holomorphic sector has exactly the same structure.

The world-sheet theory of the $M^6$ part consists of
 one free bosonic field $X^{\bar \mu}(z,{\bar z})$
 and two free fermionic fields
 $\Psi^{\bar \mu}(z)$ and ${\bar \Psi}^{\bar \mu}({\bar z})$.
In the holomorphic sector
 the energy-momentum tenor $T^{M^6}(z)$
 and the supercurrent $G^{M^6}(z)$ are
\begin{eqnarray}
 T^{M^6} &=& - {1 \over 2}
               \partial X^{\bar \mu} 
               \partial X_{\bar \mu}
             + {1 \over 2}
               \Psi^{\bar \mu} \partial \Psi_{\bar \mu},
\\
 G^{M^6} &=& \Psi^{\bar \mu} \partial X_{\bar \mu}.
\end{eqnarray}
The anti-holomorphic sector has exactly the same structure,
 and the $M^6$ part has $(N,{\tilde N}) = (1,1)$
 superconformal symmetry.
The whole world-sheet theory
 has $(N,{\tilde N}) = (1,1)$ superconformal symmetry,
 and we take
 $T=T^{M^6}+T^{W^{(4)}_k}$ and $G=G^{M^6}+G_4^{W^{(4)}_k}$
 as the currents of that symmetry.
The super-Virasoro generators $L_n$ ($n \in {\bf Z}$)
 and $G_r$ ($r \in {\bf Z}+1/2$ in Neveu-Schwarz sector
 and $r \in {\bf Z}$ in Ramond sector) are defined by
\begin{eqnarray}
 L_n &=& \oint {{dz} \over {2 \pi i z}} z^{n+2} \ T(z),
\\
 G_r &=& \oint {{dz} \over {2 \pi i z}} z^{r+3/2} \ G(z) .
\end{eqnarray}
The anti-holomorphic sector has exactly the same structure.
The partition function, or one-loop vacuum amplitude,
 is explicitly calculated in Ref.\cite{AFK}
 for even $k$.
We consider $k$ as an even number from now on. 
The difference between type IIA and type IIB theories
 is the difference of the usual universal GSO projection.
In addition to the universal GSO projection,
 we have to do the {\it additional GSO projection}
 by which the half breaking of the space-time supersymmetry
 is realized in the world-sheet theory.

\section{Boundary states for Propagators}
\label{sec:boundary}

The propagator of the closed string in type II theories
 can be calculated as the transition amplitude
 between two appropriate boundary states
 by the propagator operator
 in the world-sheet conformal field theory.
In case of the flat space-time,
 the propagator of bosonic states (in the bosonic string theory)
 is extensively studied in Ref.\cite{CMNP},
 and the propagators in the superstring theory
 are calculated in Refs.\cite{Fainberg-Marshakov,Marshakov}.
In order to obtain the propagator
 in the field theory limit (not the low-energy limit),
 the {\it point boundary states},
 which describe the states of the closed string
 shrinking to a point in space-time,
 should be used.
For the bosonic (NS-NS or R-R) states of strings,
 we can use D(-1)-brane, or D-instanton, states\cite{DFPSLR}:
\begin{equation}
 | B(y) \rangle = | B_X(y) \rangle
                \otimes | B_\psi \rangle
                \otimes  | B_{\rm gh} \rangle
                \otimes | B_{\rm sgh} \rangle,
\end{equation}
 where
\begin{equation}
 | B_X(y) \rangle
 = \delta^{10} ({\hat x}-y)
   \exp \left\{
    {\sum_{n=1}^\infty {1 \over n}
     \alpha^{\hat \mu}_{-n} {\tilde \alpha}_{-n {\hat \mu}}} \right\}
   | 0 \rangle
\label{B_X}
\end{equation}
 describes the state of the closed string
 shrinking at a space-time point of $y$ 
 (or the state where the edge of the open string
 is fixed at a space-time point of $y$),
 $| B_\psi \rangle$ is the world-sheet fermion contribution
 which is determined by the supersymmetry and T-duality,
 and $| B_{\rm gh} \rangle \otimes | B_{\rm sgh} \rangle$
 is the contribution of world-sheet ghost fields,
 the $bc$-ghost state and the $\beta\gamma$-ghost state, respectively.
(We are using the notation of Ref.\cite{Polchinski}
 except for a factor $i$ for world-sheet fermion fields.)
The propagator
\begin{equation}
 P_B(y'-y) = \langle B(y') | D_B | B(y) \rangle
\end{equation}
 with the propagator operator
\begin{equation}
 D_B = {{\alpha'} \over {4\pi}} \cdot {1 \over 2}
       \int_{|z| \le 1} d^2z {1 \over {|z|^2}}
       z^{L_0} {\bar z}^{{\tilde L}_0}
\end{equation}
 can be explicitly calculated.

For the fermionic (NS-R or R-NS) states of strings,
 we can not use the D-instanton states,
 because they do not couple with a single fermionic state.
Therefore,
 we have to introduce some source boundary states
 which couple with a single fermionic state.
The source states do not necessarily correspond to
 some physical objects like D-branes.
We concretely introduce the following boundary state
 for NS-R fermionic states.
\begin{equation}
 | B(y,{\bf s}) \rangle^{\hat \mu}
 = | B_X(y) \rangle \otimes | B_{\rm gh} \rangle
 \otimes \psi_{-1/2}^{\hat \mu} | 0 \rangle_{\rm NS}
 \otimes {\widetilde {| {\psi_{\bf s}} \rangle}}_{\rm R},
\end{equation}
 where $| 0 \rangle_{\rm NS}$ is the Neveu-Schwarz vacuum state
 and $\widetilde{ | \psi_{\bf s} \rangle}_{\rm R}$
 is the Ramond vacuum state.
The spin state
 ${\bf s} = s_1 \otimes s_2 \otimes s_3 \otimes s_4 \otimes s_5$
 with $s_i = \pm 1/2$ for $i=1,2,\cdots,5$
 is described by
\begin{equation}
 \psi_{\bf s}
  = \eta_1 \otimes \eta_2 \otimes\eta_3 \otimes \eta_4 \otimes \eta_5
\end{equation}
 with
\begin{equation}
 \eta_i
 = \left( \begin{array}{c} 1 \\ 0 \end{array} \right), \quad
   \left( \begin{array}{c} 0 \\ 1 \end{array} \right)
\end{equation}
 for $s_i=1/2, -1/2$, respectively.
The $bc$-ghost state $| B_{\rm gh} \rangle$ is explicitly given by
\begin{equation}
 | B_{\rm gh} \rangle
 = \exp\left\{
        \sum_{n=1}^\infty
         \left(
         c_{-n} {\tilde b}_{-n} - b_{-n} {\tilde c}_{-n}
         \right)
       \right\}
    {{c_0+{\tilde c}_0} \over 2}
   | \downarrow \rangle \otimes \widetilde{ | \downarrow \rangle}
\end{equation}
and
\begin{equation}
 \langle B_{\rm gh} |
 = \widetilde{ \langle \uparrow |} \otimes \langle \uparrow |
    {{{\tilde b}_0-b_0} \over 2}
        \exp
        \left\{
        \sum_{n=1}^\infty
        \left(
         {\tilde b}_n c_n - {\tilde c}_n b_n
        \right)
        \right\}.
\end{equation}

Using the propagator operator
\begin{equation}
 D_F = {{\alpha'} \over {4\pi}} \cdot {\tilde G}_0 \cdot {1 \over 2}
       \int_{|z| \le 1} d^2z {1 \over {|z|^2}}
       z^{L_0} {\bar z}^{{\tilde L}_0},
\label{prop_op_F}
\end{equation}
 we can explicitly calculate the fermion propagator as follows.
\begin{eqnarray}
 P_F(y'-y)_{{\bf s}'{\bf s}}^{{\hat \mu} {\hat \nu}}
 &=& {}^{{\hat \mu}}\langle B(y',{\bf s}') |
     D_F
     | B(y,{\bf s}) \rangle^{\hat \nu}
\nonumber\\
 &=& -i {\sqrt{\alpha'} \over 2}
     \sum_{N_X=0}^\infty d(N_X) \cdot
     \int {{d^{10} q} \over {(2\pi)^{10}}}
     {\bar \psi}_{{\bf s}'}
      q_{\hat \rho} \Gamma^{\hat \rho} \psi_{\bf s}
     \eta^{{\hat \mu} {\hat \nu}} {1 \over {q^2 + M_{N_X}^2}}
     e^{i q_{\hat \sigma} (y'-y)^{\hat \sigma}},
\label{propagator0}
\end{eqnarray}
 where $M_{N_X}^2 = 4 N_X / \alpha'$
 with the level $N_X$ of the bosonic excitations on the world-sheet,
 and $d(N_X)$ is the degeneracy of the open superstring state
 with the bosonic level $N_X$ and zero fermionic level.
In the momentum space
\begin{equation}
 P_F(q)_{{\bf s}'{\bf s}}^{{\hat \mu} {\hat \nu}}
 =  \sum_{N_X=0}^\infty d(N_X) \cdot
     {\bar \psi}_{{\bf s}'}
      q_{\hat \rho} \Gamma^{\hat \rho} \psi_{\bf s}
     {{\eta^{{\hat \mu} {\hat \nu}}} \over {q^2 + M_{N_X}^2}}.
\label{propagator0-p}
\end{equation}
This result can be understood
 as the sum of the propagators of the fields
 which satisfy some non-local field equations.
The different GSO projection for type IIA and type IIB theories
 is corresponding to take the different chirality
 for ${\bf s}'$ and ${\bf s}$.
Not all the NS-R fermionic states
 in type IIA and type IIB string theories in the flat space-time
 are included in this propagator,
 but all the NS-R massless fermionic states are included.

To do the same in the fivebrane background,
 we need the explicit expression of the D-instanton state
 for the bosonic sector
 in the fivebrane background.

The D-brane boundary state in the SU$(2)_k$ WZW model
 is discussed in Refs.\cite{Alekseev-Schomerus,FFFS}.
It is possible to explicitly construct the D-instanton state.
The place where the edge of the open string is fixed
 is the north or south pole of $S^3$.
The D-instanton state is the Cardy's state
\begin{equation}
 | C \rangle = \sum_{l=0}^k \sqrt{S_0{}^l} | l \rangle,
\label{Cardy}
\end{equation}
 where
\begin{equation}
 S_{l'}{}^l =
  \sqrt{{2 \over {k+2}}}
  \sin \left( \pi {{(l'+1)(l+1)} \over {k+2}} \right)
\end{equation}
 and $| l \rangle$ is the Ishibashi state which satisfies
\begin{equation}
 \langle l' |
  q^{{1 \over 2}
  \left( L_0^{\rm WZW} + {\tilde L}_0^{\rm WZW} \right)}
 | l \rangle
 = \delta_{l'l} q^{-S_{l,k}} \chi^{(k)}_l(q)
\end{equation}
 with $S_{l,k}=((l+1)/2)^2/(k+2)-3/24$
 and the SU$(2)_k$ character $\chi^{(k)}_l(q)$.
In the state $| C \rangle$
 there is no explicit dependence on the spatial point
 where the edge of the open string is fixed.
This is due to the fact that
 the state is constructed in the SU$(2)_k$ invariant way.

The D-brane boundary state in the linear-dilaton background
 is investigated in Refs.\cite{Li,Rajaraman-Rozali}.
It is impossible to impose Dirichlet boundary condition
 in conformal invariant way to the Feigin-Fuchs field
 because of the non-zero background charge $Q$\cite{Li}.
Although we can construct the boundary state
 which satisfies Dirichlet boundary condition
 in the same way as Eq.(\ref{B_X}),
 it is not conformal invariant.
The Dirichlet boundary state can be defined,
 not directly applying the Dirichlet boundary condition,
 but using the Cardy's condition\cite{Rajaraman-Rozali}.
The states are described in the similar way of Eq.(\ref{Cardy}):
\begin{equation}
 | B_{\rm FF}(y) \rangle
 = \int {{dp} \over {2\pi}} e^{-i p y} | p \rangle_{\rm FF},
\label{B_FF}
\end{equation}
 where
\begin{equation}
 {}_{\rm FF}\langle p' |
  q^{L_0^{\rm FF} + {\tilde L}_0^{\rm FF} - {c_{\rm FF} \over {12}}}
 | p \rangle_{\rm FF}
 = \delta (p'-p) \ q^{p^2} q^{-1/12}
   \prod_{n=1}^\infty \left( 1 - q^{2n} \right)^{-1}
\end{equation}
 with $c_{\rm FF} = 1 + 6 Q^2 / \alpha'$.
The states $| p \rangle_{\rm FF}$
 with a continuous parameter $-\infty < p < \infty$
 are the highest weight states of the continuous representation
 of the linear-dilaton conformal field theory
 with weight
 ${{\alpha'} \over 4} ( p^2 + ( Q/\alpha')^2 )$.
The boundary state $| B(y) \rangle_{\rm FF}$ satisfies
 the Cardy's condition.
\begin{equation}
 \langle B_{\rm FF}(y) |
  q^{L_0^{\rm FF} + {\tilde L}_0^{\rm FF} - {c_{\rm FF} \over {12}}}
 | B_{\rm FF}(y) \rangle
 = {\tilde q}^{-1/12}
   \prod_{n=1}^\infty \left( 1 - {\tilde q}^{2n} \right)^{-1},
\end{equation}
 where ${\tilde q} \equiv e^{-2 \pi i / \tau}$
 with $q \equiv e^{2 \pi i \tau}$.
The parameter $y$ does not necessarily mean the position
 where the edge of the open string is fixed.
But in small $Q$ limit,
 which is the limit we will consider,
 it has the meaning of the position.
Here, we note that the discrete representation of
 the linear dilaton conformal field theory need not be considered
 in the calculation of the tree-level propagator
 corresponding to the two-point amplitude at genus zero\cite{AFK}.

Using the D-instanton boundary states
 for the SU$(2)_k$ WZW model and the linear-dilaton theory,
 we introduce the following boundary state
 for the calculation of the fermion propagator.
\begin{equation}
 | B(y,{\bf s}) \rangle^{\hat \mu}
 = | B_X(y) \rangle_{M^6} \otimes | B_{\rm FF}(y) \rangle
   \otimes | C \rangle \otimes | B_{\rm gh} \rangle
   \otimes | B_\psi({\bf s}) \rangle^{\hat \mu},
\end{equation}
where $| B_\psi({\bf s}) \rangle^{\hat \mu}$
 is the contribution of the world-sheet fermions in NS-R sector.
The state $| B_\psi({\bf s}) \rangle^{\hat \mu}$
 has a little complicated form due to the additional GSO projection.
\begin{equation}
 | B_\psi({\bf s}) \rangle^{\hat \mu}
 = \left\{
   \begin{array}{ll}
    \psi^{\bar \mu}_{-1/2} | 0 \rangle_{\rm NS}
     \otimes \widetilde{| \psi_{\bf s} \rangle}_{\rm R},
   &
    \quad (s_4, s_5) = (1/2, -1/2), (-1/2, 1/2)
   \\
    \psi^\mu_{-1/2} | 0 \rangle_{\rm NS}
     \otimes \widetilde{| \psi_{\bf s} \rangle}_{\rm R},
   &
    \quad (s_4, s_5) = (1/2, 1/2), (-1/2, -1/2)
   \end{array}
   \right.
\label{f-state}
\end{equation}
 with totally odd number of $s_i=-1/2$
 (odd ten-dimensional chirality) for type IIA theory
 and with totally even number of $s_i=-1/2$
 (even ten-dimensional chirality) for type IIB theory.
The space-time index of the state
 is restricted depending on the spin state,
 and the number of the allowed states is the half of the states
 in case of the flat space-time.
This is the realization of the half supersymmetry breaking
 by the fivebrane background in the world-sheet theory\cite{AFK}.

Now,
 we can calculate the fermion propagator in the fivebrane background
 as the transition amplitude between these source boundary states
 by the propagator operator of Eq.(\ref{prop_op_F})
 with the operators $L_0$, ${\tilde L}_0$ and ${\tilde G}_0$
 defined in the previous section.

\section{Fermion propagators in fivebrane backgrounds}
\label{sec:propagator}

The explicit form of the fermion propagator is described as follows.
\begin{eqnarray}
 P_F(y'-y)_{{\bf s}'{\bf s}}^{{\hat \mu}{\hat \nu}} &=& 
 {}^{\hat \mu}\langle B(y',{\bf s}') |
  D_F | B(y,{\bf s}) \rangle^{\hat \nu}
\nonumber\\
 &=& {{\alpha'} \over {4\pi}} \cdot {1 \over 2}
     \int_{|z| \le 1} d^2 z {1 \over {|z|^2}} \
     {}^{\hat \mu}\langle B(y',{\bf s}') |
      {\tilde G}_0 z^{L_0} {\bar z}^{{\tilde L}_0}
     | B(y,{\bf s}) \rangle^{\hat \nu}.
\label{propagator1}
\end{eqnarray}
Here,
\begin{eqnarray}
 L_0 &=& L_0^{M^6} + L_0^{\rm FF} + L_0^{WZW} + L_0^{\rm ghost}
         - \nu,
\\
 {\tilde L}_0
  &=& {\tilde L}_0^{M^6} + {\tilde L}_0^{\rm FF}
    + {\tilde L}_0^{WZW} + {\tilde L}_0^{\rm ghost}
    - {\tilde \nu}
\end{eqnarray}
 with
\begin{eqnarray}
 L_0^{M^6} &=&
  {{\alpha'} \over 4} p^{\bar \mu} p_{\bar \mu}
  + \sum_{n=1}^\infty \alpha_{-n}^{\bar \mu} \alpha_{{\bar \mu} n}
  + \sum_{r+\nu=1}^\infty r \psi_{-r}^{\bar \mu} \psi_{{\bar \mu} r},
\\
 L_0^{\rm FF} &=&
  {{\alpha'} \over 4} \left( p^6 - {{iQ} \over {\alpha'}} \right)^2
  + {{\alpha'} \over 4} \left( {Q \over {\alpha'}} \right)^2
  + \sum_{n=1}^\infty \alpha^6_{-n} \alpha^6_n
  + \sum_{r+\nu=1}^\infty r \psi_{-r}^6 \psi_r^6,
\\
 L_0^{WZW} &=&
  L_0^{WZW}\vert_B
  + \sum_{r+\nu=1}^\infty r \psi_{-r}^{\mu'} \psi_r^{\mu'},
\\
 L_0^{\rm ghost} &=&
  \sum_{n=1}^\infty
   n \left( b_{-n} c_n + c_{-n} b_n \right)
  + \sum_{r+\nu=1}^\infty
   r \left( \beta_{-r} \gamma_r - \gamma_{-r} \beta_r \right),
\end{eqnarray}
 and the same for ${\tilde L}_0$.
The eigenvalue of the momentum operator $p^6-iQ/\alpha'$
 corresponds to $p$ in Eq.(\ref{B_FF}),
 $L_0^{WZW}\vert_B$ is the bosonic contribution,
 and $\mu'=7,8,9$.
For NS-R sector, $\nu=1/2$ and ${\tilde \nu}=0$.
Furthermore,
\begin{equation}
 {\tilde G}_0
 = {\tilde G}_0^{M^6} + {\tilde G}_0^{W^{(4)}_k} 
 + {\tilde G}_0^{\rm ghost}
\end{equation}
 with
\begin{eqnarray}
 {\tilde G}_0^{M^6} &=&
  - i \sqrt{{\alpha'} \over 2} p^{\bar \mu}
      {\tilde \psi}_{{\bar \mu} 0}
  + \mbox{fermion non-zero modes},
\\
 {\tilde G}_0^{W^{(4)}_k} &=&
  \sqrt{{2 \over {\alpha'}}} \sqrt{{2 \over {k+2}}}
   {\tilde J}^{\mu'}_0 {\tilde \psi}^{\mu'}_0
 -i \sqrt{{{\alpha'} \over 2}}
    \left( p^6 -{{iQ} \over {\alpha'}} \right) {\tilde \psi}^6_0
 + {1 \over {3!}} \sqrt{{2 \over {k+2}}}
   \epsilon_{\mu'\nu'\rho'} {\tilde \psi}^{\mu'}_0
                            {\tilde \psi}^{\nu'}_0
                            {\tilde \psi}^{\rho'}_0
\nonumber\\
 &+& \mbox{fermion non-zero modes},
\\
 {\tilde G}_0^{\rm ghost} &=&
 - \sum_{n=-\infty}^\infty
   \left(
    {1 \over 2} n {\tilde \beta}_{-n} {\tilde c}_n
    + 2 {\tilde b}_n {\tilde \gamma}_{-n}
   \right).
\end{eqnarray}
Here, ${\tilde J}^{\mu'}_0$ is the zero-mode
 in the mode expansion of the WZW current ${\tilde J}^{\mu'}$
 (${\tilde J}_i$ with $i=1,2,3$ in Sec.\ref{sec:CFT}
 correspond to ${\tilde J}^{\mu'}$ with $\mu'=7,8,9$, respectively).
We did not explicitly write the contributions
 of the non-zero modes of the world-sheet fermions,
 because they give no contribution to the fermion propagator
 of Eq.(\ref{propagator1}).

The calculation is straightforward except for one quantity
\begin{equation}
 \langle C | {\tilde J}_0^{\mu'} z^{L_0^{\rm WZW}\vert_B}
             {\bar z}^{{\tilde L}_0^{\rm WZW}\vert_B}
 | C \rangle.
\end{equation}
We can show that this quantity vanishes
 by using the conditions of
 conformal and SU$(2)_k$ invariance
\begin{eqnarray}
 {\tilde L}_0^{\rm WZW}\vert_B | C \rangle
  &=& L_0^{\rm WZW}\vert_B | C \rangle,
\\
 {\tilde J}_0^{\mu'} | C \rangle &=& - J_0^{\mu'} | C \rangle,
\end{eqnarray}
 and the definition of the Ishibashi state
\begin{equation}
 | l \rangle 
 = \sum_{N,j} | l; N, j \rangle
   \otimes U \widetilde {| l; N, j \rangle},
\end{equation}
 where $U$ is an unitary operator introduced in Ref.\cite{Ishibashi},
 and the states $| l; N, j \rangle$
 are in the orthonormal representations
 of the SU$(2)_k$ Ka\v c-Moody algebra with $0 \le l \le k$.
The states $| l; 0, j \rangle$ are in the orthonormal representations
 of the SU$(2)$ algebra with $-l \le j \le l$,
 and $N$ denotes the ways of operating some ${\tilde J}_{-n \mu'}$
 to the states.
We obtain
\begin{equation}
 \langle C | {\tilde J}_0^{\mu'} z^{L_0^{\rm WZW}\vert_B}
             {\bar z}^{{\tilde L}_0^{\rm WZW}\vert_B}
 | C \rangle
 = - \sum_{l,N} S_0{}^l |z|^{2 L_0(l,N)}
     \sum_j \langle l;N,j| J_0^{\mu'} |l;N,j \rangle = 0,
\end{equation}
 where $L_0(l,N)$ is the eigenvalue
 of the operator $L_0^{\rm WZW}\vert_B$
 on the state $|l;N,j \rangle$.
The second equality comes from the fact that
 the trace of $J_0^{\mu'}$ over a representation of SU$(2)$ is zero.

The result of the calculation is
\begin{eqnarray}
 P_F(y'-y)_{{\bf s}'{\bf s}}^{{\hat \mu}{\hat \nu}} &=&
 -i {\sqrt{\alpha'} \over 2}
 \int {{d^7q} \over {(2\pi)^7}}
  e^{i q_{\tilde \rho} (y'-y)^{\tilde \rho}}
  {\bar \psi}_{{\bf s}'}
   \left(
    q_{\tilde \sigma} \Gamma^{\tilde \sigma}
    + i {Q \over {\alpha'}} {1 \over {3!}}
        \epsilon_{\lambda'\kappa'\omega'}
         \Gamma^{\lambda'} \Gamma^{\kappa'} \Gamma^{\omega'}
   \right)
  \psi_{\bf s}
  \eta^{{\hat \mu}{\hat \nu}}
\nonumber\\&&
  \cdot
  \sum_{N_X=0}^\infty d(N_X)
  \sum_{l=0}^k \sum_{m=-\infty}^\infty
  \left( 2(k+2)m+l+1 \right)
  \sqrt{{2 \over {k+2}}} \sin \left( \pi {{l+1} \over {k+2}} \right)
\nonumber\\&&
  \cdot
  {1 \over
   {
   q^{\tilde \delta} q_{\tilde \delta}
   + \left( {Q \over {\alpha'}} \right)^2 + M_{N_X}^2
   + {4 \over {\alpha'}}
     \left( (k+2) m^2 + (l+1) m \right)
   }},
\label{propagator2}
\end{eqnarray}
 where ${\tilde \rho},{\tilde \sigma},{\tilde \delta}=0,1,\cdots,6$.
We used the explicit formula for SU$(2)_k$ character
\begin{equation}
 \chi^{(k)}_l(q) =
 {1 \over {\eta(\tau)^3}} \cdot 2(k+2)
 \sum_{m=\infty}^\infty \left( m + {{l+1} \over {2(k+2)}} \right)
 \exp \left(
       2 \pi i (k+2) \tau
         \left( m + {{l+1} \over {2(k+2)}} \right)^2
      \right),
\end{equation}
 where $q=e^{2 \pi i \tau}$\cite{Gepner-Witten}.
In the denominator of the last factor of Eq.(\ref{propagator2})
 the first term is the squared energy-momentum
 in the flat seven-dimensional space-time,
 the second term is the universal mass shift
 due to the linear-dilaton background
 \cite{Kiritsis-Kounnas},
 and the third term is the mass by the usual string excitations
 ($M_{N_X}^2=4N_X/\alpha'$).
The fourth term is the contribution
 which breaks the ten-dimensional space-time interpretation
 in finite $k$.

The space-time interpretation
 of $M^6 \otimes W^{(k)}_4$ in this world-sheet theory
 is possible only in large $k$ limit.
In large $k$ limit
 the contribution of $m=0$ dominates in Eq.(\ref{propagator2}).
The propagator becomes
\begin{eqnarray}
 P_F(y'-y)_{{\bf s}'{\bf s}}^{{\hat \mu}{\hat \nu}} &\rightarrow&
 -i {\sqrt{\alpha'} \over 2}
  \sum_{N_X=0}^\infty d(N_X)
 \int {{d^7q} \over {(2\pi)^7}}
  e^{i q_{\tilde \rho} (y'-y)^{\tilde \rho}}
  \sum_{l=0}^k
  \left( l+1 \right)
  \sqrt{{2 \over {k+2}}} \sin \left( \pi {{l+1} \over {k+2}} \right)
\nonumber\\&&
  \cdot
  {\bar \psi}_{{\bf s}'}
   \left(
    q_{\tilde \sigma} \Gamma^{\tilde \sigma}
    + i {Q \over {\alpha'}} {1 \over {3!}}
        \epsilon_{\lambda'\kappa'\omega'}
         \Gamma^{\lambda'} \Gamma^{\kappa'} \Gamma^{\omega'}
   \right)
  \psi_{\bf s}
  \eta^{{\hat \mu}{\hat \nu}}
  {1 \over
   {
   q^{\tilde \delta} q_{\tilde \delta}
   + \left( {Q \over {\alpha'}} \right)^2 + M_{N_X}^2
   }}.
\end{eqnarray}
The summation over $l$ is exactly calculable.
\begin{equation}
  \sum_{l=0}^k
  \left( l+1 \right)
  \sqrt{{2 \over {k+2}}} \sin \left( \pi {{l+1} \over {k+2}} \right)
  = \sqrt{{{k+2} \over 2}} \cdot
    {{\sin {{\pi} \over {k+2}}} \over {1-\cos{{\pi} \over {k+2}}}}.
\end{equation}
This factor diverges in large $k$ limit as $(\sqrt{k+2})^3$,
 and can be understood as the correspondence of
 the momentum integration in the $S^3$ space
 with the radius $\alpha'/Q=\sqrt{\alpha'(k+2)}$.
Because in the state $| C \rangle$
 there is no explicit spatial point
 where the edge of the open string is fixed in $S^3$,
 the momentum in $S^3$ is not explicitly introduced.
The propagation in $S^3$
 can not be described by using this boundary state.
We obtain the propagator in the momentum space as
\begin{equation}
 P_F(q)_{{\bf s}'{\bf s}}^{{\hat \mu}{\hat \nu}} =
  \sum_{N_X=0}^\infty d(N_X)
  {\bar \psi}_{{\bf s}'}
   \left(
    q_{\tilde \sigma} \Gamma^{\tilde \sigma}
    + i {Q \over {\alpha'}} {1 \over {3!}}
        \epsilon_{\lambda'\kappa'\omega'}
         \Gamma^{\lambda'} \Gamma^{\kappa'} \Gamma^{\omega'}
   \right)
  \psi_{\bf s}
  {{\eta^{{\hat \mu}{\hat \nu}}} \over
   {
   q^{\tilde \delta} q_{\tilde \delta}
   + \left( {Q \over {\alpha'}} \right)^2 + M_{N_X}^2
   }}.
\label{propagator3}
\end{equation}
This result has the similar form to
 the propagator in the flat space-time, Eq.(\ref{propagator0-p}),
 but it includes some additional contribution
 which is proportional to three gamma matrices.
It would be interesting
 if the term could result some interesting phenomena
 like fermion pair condensation,
 but it is not the case.
We have to do further the additional GSO projection.
As can be seen in Eq.(\ref{f-state}),
 the additional GSO projection on the spinor index
 can be described as the six-dimensional chirality projection.
The polarization spinor $\psi_{\bf s}$
 should satisfy appropriate chirality conditions.
The final result can be described as follows
 not explicitly including $\psi_{\bf s}$,
 but using projection operators.
\begin{equation}
 P_F(q)^{{\hat \mu}{\hat \nu}} =
 \left\{
 \begin{array}{l}
  \displaystyle{\sum_{N_X=0}^\infty} d(N_X) \cdot
   {{1 \pm \Gamma_{11}} \over 2} \cdot
   {{1 \mp \gamma_6} \over 2} \cdot
  {{q_{\bar \sigma} \Gamma^{\bar \sigma} \eta^{{\bar \mu}{\bar \nu}}}
   \over
   {
   q^{\tilde \delta} q_{\tilde \delta}
   + \left( {Q \over {\alpha'}} \right)^2 + M_{N_X}^2
   }} \cdot
   {{1 \pm \gamma_6} \over 2} \cdot
   {{1 \mp \Gamma_{11}} \over 2}
 \\
  \displaystyle{\sum_{N_X=0}^\infty} d(N_X) \cdot
   {{1 \pm \Gamma_{11}} \over 2} \cdot
   {{1 \pm \gamma_6} \over 2} \cdot
  {{q_{\bar \sigma} \Gamma^{\bar \sigma} \eta^{{\mu}{\nu}}}
   \over
   {
   q^{\tilde \delta} q_{\tilde \delta}
   + \left( {Q \over {\alpha'}} \right)^2 + M_{N_X}^2
   }} \cdot
   {{1 \mp \gamma_6} \over 2} \cdot
   {{1 \mp \Gamma_{11}} \over 2}
 \end{array}
 \right.
\label{propagator-final}
\end{equation}
 for type IIA and type IIB theories, respectively in sign,
 where $\gamma_6 \equiv \Gamma^0 \Gamma^1 \cdots \Gamma^5$
 and $\Gamma_{11} = \Gamma^0 \Gamma^1 \cdots \Gamma^9$.

The lightest ($N_X=0$) propagation modes can be understood
 as the components of the supermultiplets
 in six-dimensional $N=(2,0)$ and $N=(1,1)$ supergravity theories
 for type IIA and type IIB theories, respectively.
Note that there are no propagators of the fields
 in SO$(4)$ (local Lorentz) spinor representations
 with $\Gamma^\mu$ ($\mu=6,7,8,9$) in the numerator,
 which is consistent with the arguments on the fermion zero-modes
 in the fivebrane background of low-energy supergravity theories.
Namely,
 the existence of four fermion zero-modes
 in non-trivial four-dimensional space $W^{(4)}_k$ means
 that there are no propagators (or two point functions)
 in SO$(4)$ spinor representations,
 if we believe the path integral formalism for quantization.

In case of type IIA (IIB) theory and $N_X=0$,
 the propagator of the first line of Eq.(\ref{propagator-final})
 corresponds to the propagations of
 one complex six-dimensional spin-3/2 gravitino
 and one complex six-dimensional spin-1/2 dilatino
 in the $N=(2,0)$ ($N=(1,1)$) supergravity multiplet.
Another pair of the gravitino and dilatino
 in the supergravity multiplet
 should be supplied from the R-NS sector.
The propagator of the second line of Eq.(\ref{propagator-final})
 corresponds to the propagations of four complex spin-1/2 spinor fields
 in two $N=(2,0)$ tensor multiplets (two $N=(1,1)$ vector multiplets)
 of the supergravity theory.
The R-NS sector gives
 the fermion components of other two $N=(2,0)$ tensor multiplets
 (two $N=(1,1)$ vector multiplets).

Note that the obtained propagator
 correctly describes the propagation
 at around a certain distance from the place of the fivebrane.
The magnitude of the value of the dilaton field
 becomes large near and far away from the place of the fivebrane
 (see Eq.(\ref{dilaton}) with $e^{2\Phi_0}=0$ and $n=k+2$),
 and the perturbative calculation on the string coupling
 (the genus expansion) does not give correct results.
Therefore,
 the six-dimensional system mentioned in the previous paragraph
 is not the one confined in the fivebrane,
 but the system of the bulk with the distance around
 $\sqrt{\alpha'(k+2)}$ away from the place of the fivebrane.

\section{Conclusions}
\label{sec:conclusions}

We have calculated the fermion propagators
 in the fivebrane background of type IIA and type IIB string theories.
The world-sheet conformal field theory is the combination of
 the SU$(2)_k$ WZW model and linear-dilaton theory,
 which describes the non-trivial four-dimensional space,
 and the theory corresponding to the six-dimensional
 flat Minkowski space-time.
The Ka\v c-Moody level $k$
 has been taken even and large number
 for the ten-dimensional space-time interpretation.
The tree-level calculation is effective
 at the distance around $\sqrt{\alpha'(k+2)}$ away
 from the place of the fivebrane,
 where the magnitude of the value of the dilaton field is small.

The form of the obtained propagator is simple,
 and the lightest propagating modes can be understood
 as the components of the supermultiplets in six-dimensional
 $N=2$ supergravity theories.
The fact that
 there are no propagations of the fields
 in SO$(4)$ (local Lorentz) spinor representations
 is consistent with the arguments on the fermion zero-modes
 in low-energy supergravity theories.
No signature of the fermion pair condensation has found,
 which is also consistent with the arguments on the fermion zero-modes
 in low-energy effective theories.

It might be interesting
 to attempt the similar calculation in the heterotic string theory.
From the arguments
 on the fermion zero-modes in the low-energy supergravity theory,
 the pair condensations of gravitino, dilatino and gaugino
 are expected.
Although it is difficult to construct
 the source boundary state for gravitino and dilatino
 without destroying Dirichlet boundary condition,
 it is possible to construct the appropriate boundary state
 for gaugino.
The formation of the gaugino condensation
 in the low-energy supergravity theory can be understood
 as the instanton effect at the semi-classical level.
It might be interesting to ask the question
 how the semi-classical effect in the low-energy effective theory
 is described in the string world-sheet theory. 

\acknowledgments

I would like to thank
 S.~Saito and S.V.~Ketov for useful comments.
I would specially thank to B.-Y.~Hou and R.-H.~Yue
 for valuable discussions and kind hospitality
 during my stay in Xibei university.


\begin{references}
\bibitem{Witten}
 E.~Witten, Nucl. Phys. B188, 513 (1981);
 Commun. Math. Phys. 100, 197 (1985).
\bibitem{KMP}
 K.~Konishi, N.~Magnoli and H.~Panagopoulos,
 Nucl. Phys. B309, 201 (1988); B323, 441 (1989).
\bibitem{Giddings-Strominger}
 S.B.~Giddings and A.~Strominger, Nucl. Phys. B306, 890 (1988).
\bibitem{Coleman}
 S.~Coleman, Nucl. Phys. B307, 867 (1988).
\bibitem{Rey1}
 S.-J.~Rey, UCSB-TH-89/49,
 Invited talk given at Workshop on Superstrings and Particle Theory,
 Tuscaloosa, Alabama, Nov 8-11, 1989;
 Phys. Rev. D43, 526 (1991).
\bibitem{CHS1}
 C.G.~Callan, J.A.~Harvey and A.~Strominger,
 Nucl. Phys. B359, 611 (1991).
\bibitem{CHS2}
 C.G.~Callan, J.A.~Harvey and A.~Strominger,
 Nucl. Phys. B367, 60 (1991).
\bibitem{Rey2}
 S.-J.~Rey, SLAC-PUB-5659,
 Presented at Particle and Fields '91 Conf.,
 Vancouver, Canada, Aug 18-22, 1991.
\bibitem{AFK}
 I.~Antoniadis, S.~Ferrara and C.~Kounnas,
 Nucl. Phys. B421, 343 (1994).
\bibitem{Fainberg-Marshakov}
 V.Ya.~Fainberg and A.V.~Marshakov, Phys. Lett. B211, 81 (1988).
\bibitem{Marshakov}
 A.V.~Marshakov, Nucl. Phys. B312, 178 (1989).
\bibitem{Kitazawa1}
 N.~Kitazawa, Mod. Phys. Lett. A17, 2617 (2002).
\bibitem{Katagiri-Kitazawa}
 Y.~Katagiri and N.~Kitazawa, hep-th/0208069.
\bibitem{CMNP}
 A.~Cohen, G.~Moore, P.~Nelson and J.~Polchinski,
 Nucl. Phys. B267, 143 (1986).
\bibitem{DFPSLR}
 P.~Di Vecchia, M.~Frau, I.~Pesando, S.~Sciuto, A.~Lerda and R.~Russo,
 Nucl. Phys. B507, 259 (2997).
\bibitem{Polchinski}
 J.~Polchinski, ``String Theory'' Vol. I and II,
 Cambridge Univ. Pr. (1998).
\bibitem{Alekseev-Schomerus}
 A.~Yu.~Alekseev and V.~Schomerus, Phys. Rev. D60, 061901 (1999).
\bibitem{FFFS}
 G.~Felder, J.~Fr\"ohlich, J.~Fuchs and C.~Schweigert,
 J. Geom. Phys. 34 (2000) 162.
\bibitem{Li}
 M.~Li, Phys. Rev. D54, 1644 (1996).
\bibitem{Rajaraman-Rozali}
 A.~Rajaraman and M.~Rozali, JHEP 9912:005 (1999).
\bibitem{Ishibashi}
 N.~Ishibashi, Mod. Phys. Lett. A4, 251 (1989).
\bibitem{Gepner-Witten}
 D.~Gepner and E.~Witten, Nucl. Phys. B278, 493 (1986).
\bibitem{Kiritsis-Kounnas}
 E.~Kiritsis and C.~Kounnas,
 Nucl. Phys. B456, 699 (1995); Nucl. Phys. B422, 472 (1995).
\end{references}
\end{document}